\documentclass[final,3p,times,twocolumn]{elsarticle}

\usepackage{graphicx,stfloats}
\usepackage{color}
\usepackage{amsmath,amssymb}
\usepackage{tikz}
\usepackage{upgreek}
\usepackage[symbol*]{footmisc}
\graphicspath{{./figs/}}

\newcommand{\vect}[1]{\boldsymbol{#1}}
\newcommand{\pfr}[2]{\ensuremath{\frac{\partial #1}{\partial #2}}}

\newcommand{\erfc}{\mathrm{erfc}}
\newcommand{\ee}{\mathrm{e}}

\newcommand{\dd}{\mathrm{d}}
\newcommand{\YFh}{\hat{Y}_{\scriptscriptstyle {\rm F}}}
\newcommand{\YOh}{\hat{Y}_{\scriptscriptstyle {\rm O}}}
\newcommand{\Th}{\hat{T}}
\newcommand{\TFh}{\hat{T}_F}
\newcommand{\Tsh}{\hat{T}_s}
\newcommand{\Bh}{\hat{B}}
\newcommand{\oh}{\hat{\omega}}
\newcommand{\M}{\mathcal{M}}
\newcommand{\bes}{\beta_s}
\newcommand{\sis}{\sigma_s}
\newcommand{\ep}{\epsilon}
\newcommand{\Ut}{\tilde U}
\newcommand{\St}{\tilde S}
\newcommand{\At}{\tilde A}

\biboptions{sort&compress}

\journal{Proceedings of the Combustion Institute}

\begin{document}

\begin{frontmatter}

\title{Influences of stoichiometry on steadily propagating triple flames in counterflows}

\author{Prabakaran Rajamanickam\corref{cor1}}
\ead{prajaman@ucsd.edu}

\author{Wilfried Coenen}
\author{Antonio L. S\'anchez}
\author{Forman A. Williams}

\address{Department of Mechanical and Aerospace Engineering, 
          University of California San Diego, La Jolla, CA 92093--0411, USA}

\cortext[cor1]{Corresponding author:}

\begin{abstract}

Most studies of triple flames in counterflowing streams of fuel and oxidizer have been focused on the symmetric problem in which the stoichiometric mixture fraction is $1/2$. There then exist lean and rich premixed flames of roughly equal strengths, with a diffusion flame trailing behind from the stoichiometric point at which they meet. In the majority of realistic situations, however, the stoichiometric mixture fraction departs appreciably from unity, typically being quite small. With the objective of clarifying the influences of stoichiometry, attention is focused on one of the simplest possible models, addressed here mainly by numerical integration.  When the stoichiometric mixture fraction departs appreciably from $1/2$, one of the premixed wings is found to be dominant to such an extent that the diffusion flame and the other premixed flame are very weak by comparison. These curved, partially premixed flames are expected to be relevant in realistic configurations. In addition, a simple kinematic balance is shown to predict the shape of the front and the propagation velocity reasonably well in the limit of low stretch and low curvature. 
\end{abstract}


\end{frontmatter}
\renewcommand{\thefootnote}{\fnsymbol{footnote}}
\section{Introduction}
\label{sec:intro}

Triple flames, first identified by Phillips over fifty years ago~\cite{Phillips1975}, play a fundamental role in many practical combustion systems. Since they have been observed to move along mixing layers that are strained in laminar and turbulent jet flows, for example, there is interest in investigating their response to strain. The counterflow mixing layer separating two opposed planar jets of fuel and oxidizer, used in previously in theoretical~\cite{Daou19981,Daou19982} and  experimental~\cite{Ronney2006,Ronney2017} studies, provides an attractive canonical problem for analyzing these effects. 

The present contribution is intended to offer some clarifications concerning such steadily propagating triple flames, by building on a simplification of the formulation of Daou and Li{\~n}\'an~\cite{Daou19981,Daou19982}, who emphasized effects of Lewis numbers by parametrically studying, both numerically and analytically, these triple flames in mixtures with unequal diffusivities. Since underlying influences of stoichiometry tend to be obscured by varying Lewis numbers, the present considerations are restricted to equi-diffusional systems in which all Lewis numbers are unity. Effects of variable-densities~\cite{Rogg2005,Matalon2015} and heterogeneous mixtures~\cite{sivashinsky2012} introduce a number of additional interesting phenomena, but are not considered here because the emphasis is on other aspect of the problem that can be addressed more clearly without introducing these complications. Under these restrictions, implications are considered here for systems with stoichiometries that are quite likely to be encountered in practice. The simplifications that will be introduced in the formulation will be identical to those in these previous references~\cite{Daou19981,Daou19982}, simplifications which also are employed in a number of other publications~\cite{Linan1974,Dold1989,Dold1991,Dold19912,revuelta2002laminar}, thereby facilitating comparisons. 

\section{Formulation}
\label{sec:form}

\begin{figure}[ht]
\centering
\includegraphics[scale=0.35]{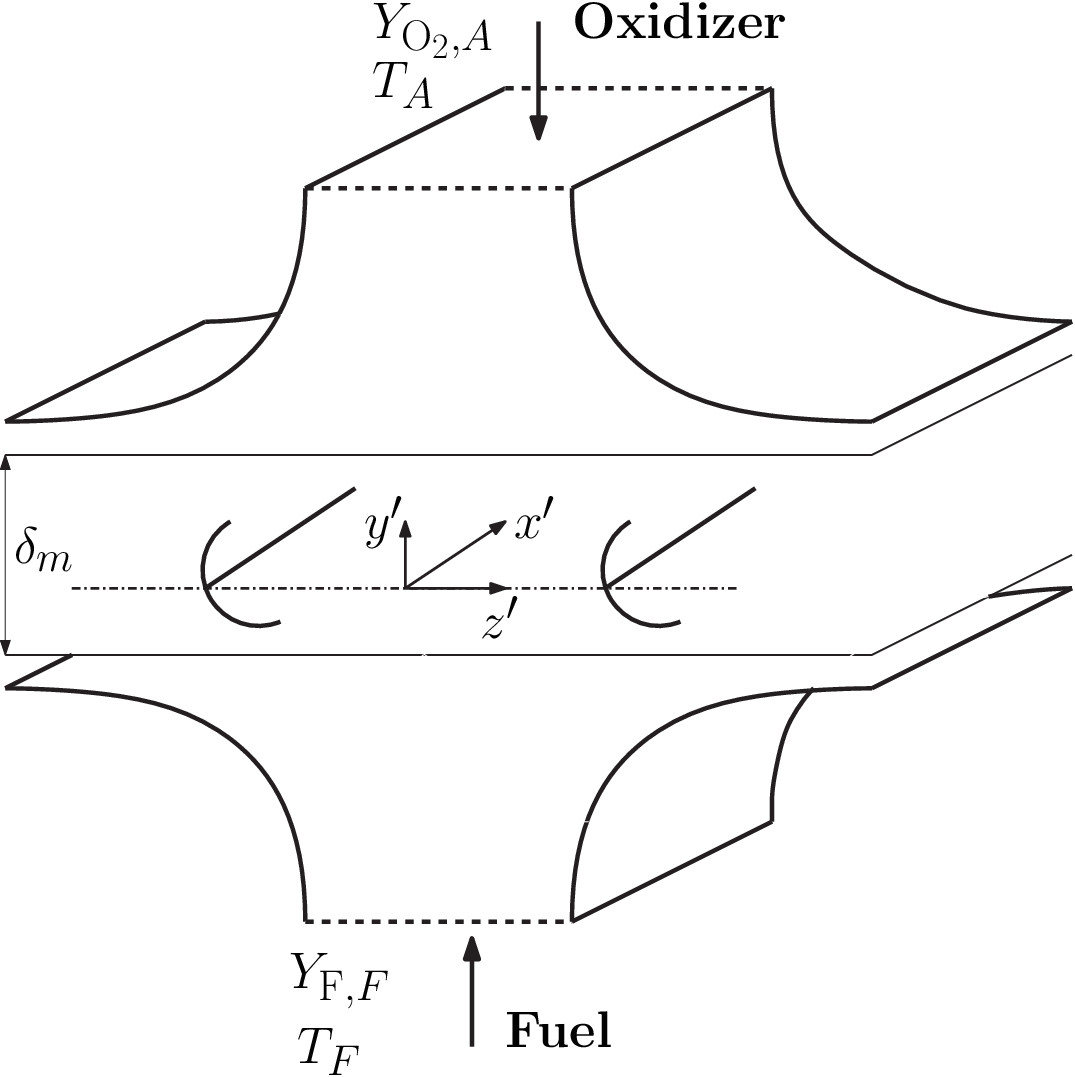}
\caption{A Schematic diagram  of the counterflow mixing layer considered.}
\label{fig:scheme}
\end{figure}

The analysis adopts a one-step irreversible reaction for the chemistry, one unit mass of fuel reacting with $s$ units of mass of oxygen to generate products, according to $\mathrm{F} + s \mathrm{O_2} \rightarrow (1+s) \mathrm{P} + q$, where $q$ denotes the amount of energy released in the process per unit mass of fuel consumed. The number of moles of fuel burned per unit volume per unit time,
\begin{equation}
    \omega = B \left(\frac{\rho Y_{\scriptscriptstyle {\rm F}}}{W_{\scriptscriptstyle {\rm F}}}\right) \left(\frac{\rho Y_{\scriptscriptstyle {\rm O_2}}}{W_{\scriptscriptstyle {\rm O_2}}}\right) \ee^{-E_a/RT}, \label{eq:omega}
\end{equation}
involves a pre-exponential factor $B$ and an activation energy $E_a$. Here, $\rho $ and  $T$ are the density and  temperature of the gas mixture, and $R$ is the universal gas constant. Mass fractions and molecular weights of species $i$ are represented by $Y_i$ and $W_i$, respectively. Following~\cite{Daou19982}, we consider a strained mixing layer configuration as shown in figure~\ref{fig:scheme}, with the front propagating at a constant speed $U$ in the negative $x'$ direction. To render the problem steady, a reference frame moving with the front will be used in the description, with the counterflowing streams approaching from $y'=\pm\infty$ and leaving at $z'=\pm\infty$. 

In the thermo-diffusive approximation (i.e. constant density and constant transport properties), the counterflow velocity field reduces to the familiar stagnation-point solution $(v,w)=(-Ay',Az')$ in terms of the strain rate $A$, which defines, together with the thermal diffusivity $D_T$, the characteristic mixing-layer thickness $\delta_m = (D_T/A)^{1/2}$. Although the velocity varies in the $z'$ direction, the temperature and composition fields are independent of $z'$ in the configuration considered. 

With the dimensionless variables
\begin{equation}
    x = \frac{x'}{\delta_m}, \ y = \frac{y'}{\delta_m}, \ \YFh = \frac{Y_{\scriptscriptstyle {\rm F}}}{Y_{\scriptscriptstyle {\rm F},F}}, \ \YOh = \frac{Y_{\scriptscriptstyle {\rm O_2}}}{Y_{\scriptscriptstyle {\rm O_2},A}}, \ \Th = \frac{T-T_A}{\gamma T_A},
\end{equation} 
and the stoichiometric mass ratio $S$ (i.e. the amount of oxygen needed to burn the unit mass of the fuel stream completely), the non-dimensional heat release $\gamma$, and the reciprocal-time pre-exponential factor $\hat B$, namely
\begin{equation}
    S=\frac{sY_{\scriptscriptstyle {\rm F},F}}{Y_{\scriptscriptstyle {\rm O_2},A}}, \ \gamma = \frac{qY_{\scriptscriptstyle {\rm F},F}}{c_pT_A(1+S)}, \ \Bh = \frac{\rho B Y_{\scriptscriptstyle {\rm O_2},A}}{W_{\scriptscriptstyle {\rm O_2}}}
\end{equation}
(where $c_p$ is the specific heat at constant pressure), the relevant problem is to solve for the temperature field and obtain the concentrations of reactants through a mixture fraction, defined as
\begin{equation}
    Z = \frac{1}{2} \erfc \left(\frac{y}{\sqrt 2}\right)=\frac{S\YFh -\YOh+1}{{S+1}} = \frac{\Th + (1+S) \YFh}{\TFh + 1+ S}. \label{Zdef}
\end{equation}
where $\TFh=(T_F-T_A)/\gamma T_A$ measures the difference in temperature between the fuel and oxidizer streams. The governing equation then becomes\footnote[2]{Here, $\Ut=U/S_{L\infty,s}$ and $\At=D_T A/S_{L\infty,s}^2$, where $ S_{L\infty,s} = \left[4 (1-Z_s)\bes^{-3}\hat B D_T \ee^{-E_a/RT_s}\right]^{1/2}$ is the stoichiometric planar velocity obtained at leading order in the limit $\bes\gg 1$, with $T_s= (1+\gamma) T_A + (T_F-T_A)Z_s$, $\bes= \frac{E_a}{RT_s}\frac{T_s-T_A}{T_s}$ and $Z_s=1/(S+1)$ being, respectively, the stoichiometric values of the adiabatic flame temperature, Zel'dovich number and mixture fraction.}
\begin{equation}
\frac{\Ut}{\sqrt{\At}} \pfr{\Th}{x} -y \pfr{\Th}{y} = \pfr{^2\Th}{x^2} + \pfr{^2\Th}{y^2} + \frac{\oh}{Z_s \At}, \label{eq:finaleq}
\end{equation}
where
\begin{equation}
\oh  = \frac{\bes^3 \YFh \YOh}{4(1-Z_s)}\exp\left[-\frac{\bes(\Tsh-\Th)}{1-\sis(\Tsh-\Th)}\right] \label{eq:reactrate}
\end{equation}
is the dimensionless reaction rate, with  $\sis=\gamma T_A/T_s$ and $\Tsh=1+\TFh Z_s$, to be integrated with boundary conditions in the transverse direction,
\begin{align}
y\rightarrow -\infty: \quad \Th = \TFh, \quad y\rightarrow\infty: \quad \Th=0,
\end{align}
along with a chemically frozen upstream mixture and an emerging downstream diffusion flame, corresponding to
\begin{equation}
    x\rightarrow -\infty: \quad \Th = \TFh Z, \quad x\rightarrow\infty: \quad \pfr{\Th}{x}=0, \label{eq:bc5}
\end{equation}
the value of $\Ut$ serving as an eigenvalue that enables~\eqref{eq:bc5} to be satisfied. The reaction rate~\eqref{eq:reactrate} must be evaluated with use of $\YFh = Z \Tsh - Z_s \Th$ and $\YOh = (1-Z) + (1-Z_s) (Z\TFh -\Th)$, obtained from~\eqref{Zdef}.

\section{Numerical results}

 Since the problem defined above exhibits invariance under translations in the $x$ direction, to anchor the flame the additional condition $\Th=0.3$ is imposed at $x=0$ along the stoichiometric line $y=y_s$. The parametric values $\bes=8\ \& \ 20$, representative of the range of overall activation energies usually encountered, $\sis=0.85$ (corresponding to typical amounts of heat release in flames) and $\TFh=0$ (equal feed temperatures) are used in the integrations for three different values of stoichiometric ratio $S=(1,4,17.2)$. Here $S=4$ and $S=17.2$ are selected as representative of  the conditions found in methane-oxygen and methane-air combustion, respectively. 
 \begin{figure}[ht]
\centering
\includegraphics[width=67mm]{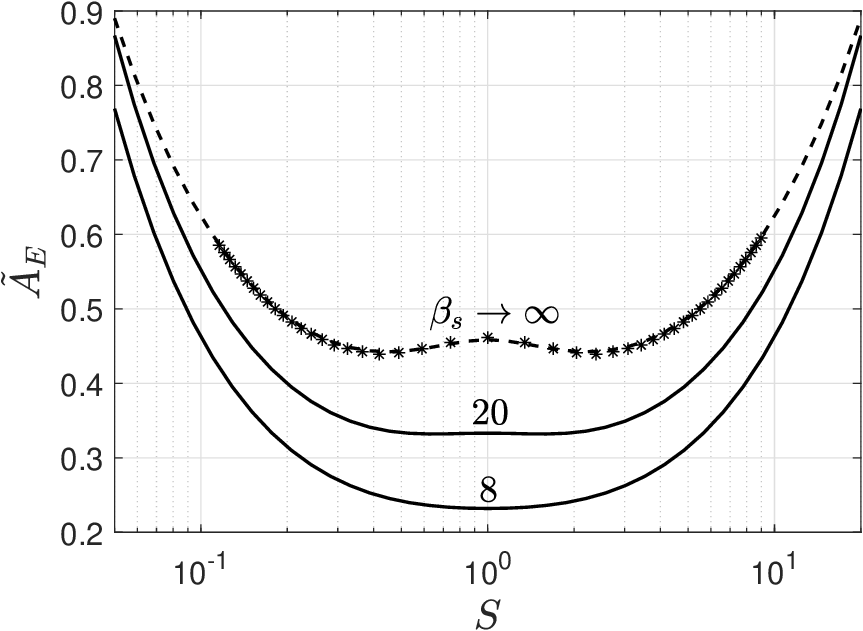}
\caption{Extinction strain rate $\At_{E}$ as a function of stoichiometric ratio $S$ for $\bes=8,\ 20,\ \sis=0.85$ and $\TFh=0$. The solid lines are obtained numerically from one-dimensional diffusion flame. The dashed line is from~\cite{Linan1974} and the points are from~\cite{Linan2017}. }   \label{fig:extinction}
\end{figure}

 Since the extinction strain rate $A_{E}$ for the one-dimensional trailing diffusion flame is of order $S_{L\infty,s}^2/D_T$, the solution for the triple flame can be anticipated to exist only for values of $\At$ in the range $0 < \At < \At_{E}\sim 1$.  The value of $\At_{E}$ is shown in figure~\ref{fig:extinction} as a function of $S$ and compared with the asymptotic predictions for $\bes \gg 1$~\cite{Linan1974,Linan2017}. In the limit $\bes\gg 1$, the curves exhibit two inflection points, one for $S>1$ and another for $S<1$. These inflection points are present in both the numerical computation~\cite{Linan2017} and the correlation formula~\cite{Linan1974}, which gives a zero slope at $S=1$, however, disappear at realistic values of $\bes$, the decrease in the temperature sensitivity of the reaction rate reversing the dependence on $S$ found in the diffusion-flame regime. 

\begin{figure*}
\centering
\includegraphics[scale=0.35]{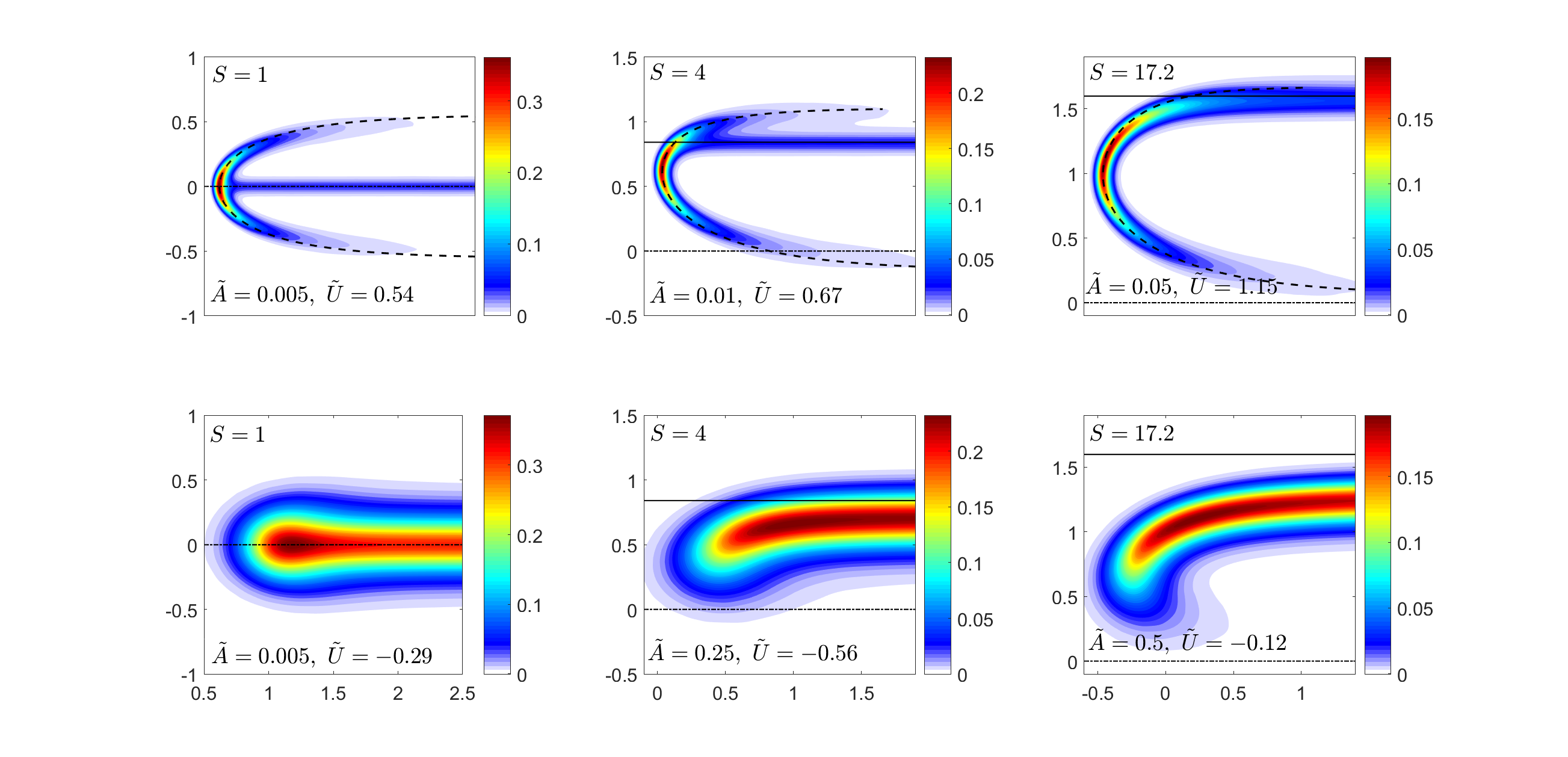}
\caption{Reaction-rate $\oh$ contours for fixed values of $\bes=8, \ \sis=0.85$ and $\TFh=0$ and for three different values of $S$ at different strain rates $\At$. Upper plots correspond to advancing fronts and lower plots show retreating fronts. Solid lines represent stoichiometric locations, and the dash-dotted line is the stagnation plane. The dashed curves are flame shapes calculated from the thin-flame analysis.}   \label{fig:reactrate}
\end{figure*}

The influence of the stoichiometry of the fuel stream on the structure of the propagating flame is investigated in figure~\ref{fig:reactrate} for $\bes=8$ by exhibiting contours of reaction rates $\oh$ defined in~\eqref{eq:reactrate}. The front shapes for $\bes=20$, not shown here, were found to be quite similar to those for $\bes=8$, except for an overall reduction in the spatial extent of the reaction region, consistent with the stronger temperature sensitivity associated with the increase in activation energy. 

To better identify the relative position of the flame, the stagnation plane $y=0$ and the stoichiometric plane $y=y_s$ are represented in each plot by a dot-dashed line and a solid line, respectively. Two values of the strain rate are selected in the figure for each value of $S$, with the smaller value on the top corresponding to an advancing front with  $\Ut>0$ and the higher value on the bottom corresponding to a retreating front with $\Ut<0$. 

The symmetric solutions for $S=1$ result in a triple-flame structure for low strain rates and a retreating edge-flame structure for near-extinction strain rates, as is well known. It can be seen, however, that the symmetric character is lost for $S=4$,  with the flame migrating to the oxidizer side of the mixing layer and the associated lean flame that develops for $y>y_s$ becoming very weak. At the higher strain rate for this value of $S$, the retreating edge flame bends away from stoichiometry, towards the stagnation plane, as it broadens.

\begin{figure}[ht]
\centering
\includegraphics[width=67mm]{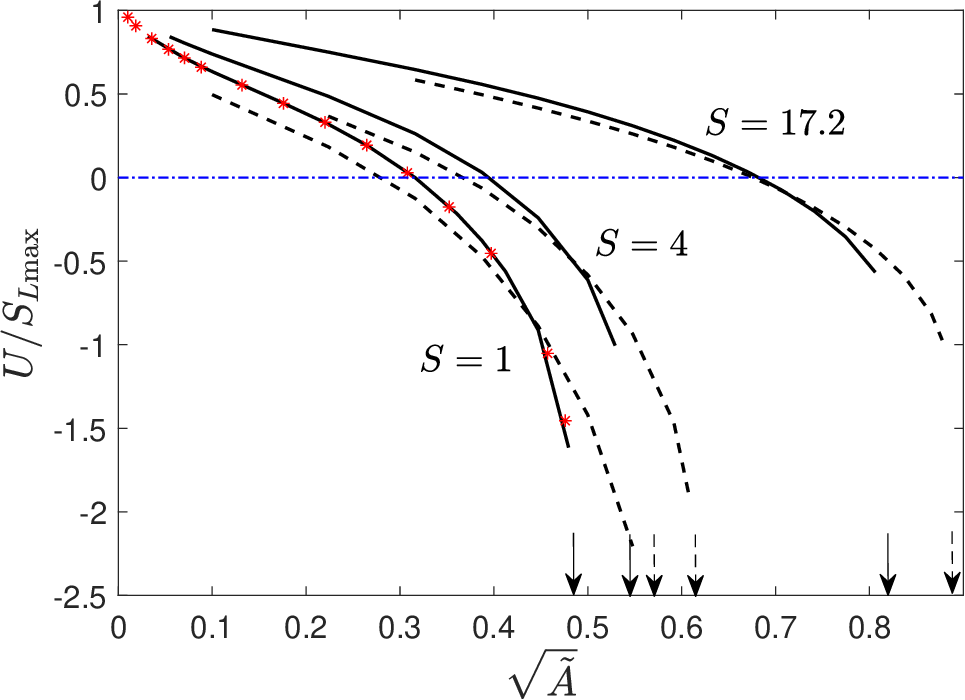}
\caption{Triple-flame propagation velocity $U$ as a function of strain rate $\At$ for $S=1,\ 4, \ 17.2 $, $\sis=0.85$ and $\TFh=0$. Solid curves correspond to $\bes=8$ and dashed curves to $\bes=20$. The asterisk marks for $S=1$ are from~\cite{Daou19982}. } \label{fig:eigenvalue}
\end{figure}

The fading lean branch disappears altogether for $S=17.2$, at which value the propagating front takes on a C shape, with one of the wings of the premixed front evolving into the trailing diffusion flame as $x \rightarrow \infty$. Also of interest is that at the lower strain rate selected for this figure, the front at $S=17.2$ is found to propagate at a velocity $\Ut=1.15>1$, that is, higher than its stoichiometric value. This behavior, already reported without explanation for the range $\At\ll \bes^{-2}$~\cite{Daou19982} through asymptotic analysis, can be explained by investigating the composition dependence of one-dimensional planar flames~\cite{praba2017}, where it is shown that the peak of the laminar planar burning velocity for these large values of $S$, in general does not lie at the stoichiometric point, but at the fuel-rich conditions. 

Although the density decrease across curved flames is well known to increase propagation velocities in configurations such as this, that influence is absent in the present constant-density analysis, leading to the higher speed arising from the effect of the planar burning velocity. While the density-change influence would be largest at stoichiometric conditions, it is seen in the left-hand figure that, on the contrary, this C-shaped flame lies entirely in a fuel-rich region, bounded by the stoichiometric and the stagnation plane. From the right-hand plot, the retreating front is seen in the figure at this value of $S\gg 1$ to become hook-shaped, with the reaction rate of the retreating edge flame actually beginning to increase very far from stoichiometry, near the stagnation plane, where the available residence times are longer, allowing the heating of the mixture by the diffusion flame to have had more time to increase the reaction rate.

The computed dependence of the flame velocity on the strain rate is shown in figure~\ref{fig:eigenvalue} for the same three different values of $S$. Since for a given activation energy $\bes$, in the thermo-diffusive approximation the two-dimensional flame propagation velocity cannot exceed its one-dimensional planar maximum speed, $S_{L,\rm{max}}$, for any strain rate $\At$ between ignition and extinction, i.e, $\Ut < \St_{L,\rm{max}}$, where $\St_{L,\rm{max}}$ is the maximum value of $\St_L=S_L/S_{L\infty,s}$, it is appropriate to plot the two-dimensional flame velocity normalized by its one-dimensional maximum velocity calculated in~\cite{praba2017}. As can be seen, for each value of $S$ there exists an $\At$ at which $\Ut=0$, thereby defining the boundary between advancing fronts and retreating fronts. The magnitude of the negative value of the propagation velocity of the retreating front goes to infinity as the strain rate approaches the extinction value $\At_{E}$, although the computations have not been carried far into the range $\Ut<0$, where convergence difficulties become more accute. The limiting strain rate is different for the solid and dashed curves, consistent with the results shown in figure~\ref{fig:extinction} for the two different activation energies considered here, the limiting values being indicated by vertical arrows at the bottom of the figure.

\section{Kinematics of thin fronts}

\label{kinematics}

As the strain rate becomes small, the flame becomes thin compared with its radius of curvature, enabling a more general analysis to be developed that is not necessarily restricted to the chemical kinetics. The initial analytical description of flame-front structures and propagation velocities, corresponding to low-strain feed streams in the present configuration, is due to Dold \text{\textit{et al.\ }}~\cite{Dold1989,Dold1991,Dold19912}, later extended to fuels with non-unity Lewis numbers by Daou \& Li{\~n}\'an~\cite{Daou19981,Daou19982}, both works invoking activation-energy asymptotics in the description. It is, however, not necessary to adopt that approach in addressing the thin-flame limit, which may be analyzed directly by treating $\ep=\sqrt{\tilde A}$ as a small parameter of expansion, thereby admitting reactions with more complex chemistry. 

A front propagating at velocity $V_f$ into a fluid whose velocity field is $\vect v$ may be described in a level-set approach by any constant value of a continuous and differentiable field function $G$ that obeys the equation 
\begin{equation}
   \vect v \cdot \nabla G = V_f |\nabla G|,
\end{equation}
when $\vect n=-\nabla G/|\nabla G|$ is the local unit vector in the direction of propagation. To apply this description to the present problem, the components of $\vect v$ are taken to be $(u,v)=(U,-Ay')$, and the field function is selected to be $G(x,y)=x-f(y)$ with $G=0$ along the front. The flame shape is then given by $x=f(y)$, conditions along the flame sheet being treated as functions of $y$.  A similar type of kinematic balance has been used previously in computing shapes of lifted flames in axisymmetric fuel jets under the additional approximation of negligible front-curvature effects \cite{revuelta2002laminar}.

In terms of planar adiabatic laminar burning velocity $\St_L(y)$ (which can be obtained irrespective of the functional form of the reaction rate), a laminar-flame thickness for a low strain rate can be defined as $\delta_L(y)/\delta_m = \sqrt{\tilde A}/\St_L$. For fronts with small curvature, the local burning velocity can be expressed in dimensional form~\cite{clavin2011,clavin2016combustion} as
\begin{equation}
    V_f = S_L - S_L \M \delta_L \kappa + \M' \delta_L \vect n\cdot \nabla \vect v \cdot \vect n. \label{eq:burn}
\end{equation}
Here $\M$ is the Markstein number for curvature and $\M'$ that for strain, which, in general, vary with $y$ along the front, $\kappa = \nabla\cdot \vect n$ is the front curvature, and $-\vect n \cdot \nabla \vect v \cdot \vect n$ being the imposed strain rate associated with the velocity gradients. 

In terms of the unknown function $g(y)=df/dy$, the non-dimensional equation then simplifies to the first-order ordinary differential equation
\begin{equation}
    \Ut + \ep yg = \St_L\sqrt{1+g^2} - \ep  \frac{\M \dd g/\dd y}{1+g^2} - \ep^2 \frac{\M' g^2}{\tilde S_L \sqrt{1+g^2}}. \label{eq:G}
\end{equation}
 The solution to this equation for $g$ describes a C-shaped flame with $g$ approaching positive infinity at $y=y_{\infty}$ and negative infinity at $y=y_{-\infty}$. This then defines a two-point boundary-value problem that possesses a continuous and differentiable solution $g(y)$ only for a particular value of the constant $\Ut$, thus constituting a nonlinear eigenvalue problem with boundary conditions,
\begin{equation}
    g\rightarrow \frac{\Ut}{\pm(\St_L-\ep^2\M'/\St_L)-\ep y}, \label{eq:Gbc}
\end{equation}
the upper sign applying at $y_{\infty}$ and the lower sign at $y_{-\infty}$, obtained as limiting forms of~\eqref{eq:G}. The domain interval $(y_{-\infty},y_\infty)$ itself is derived by setting the denominator of~\eqref{eq:Gbc} to zero.
 \begin{figure}[ht]
\centering
\includegraphics[width=67mm]{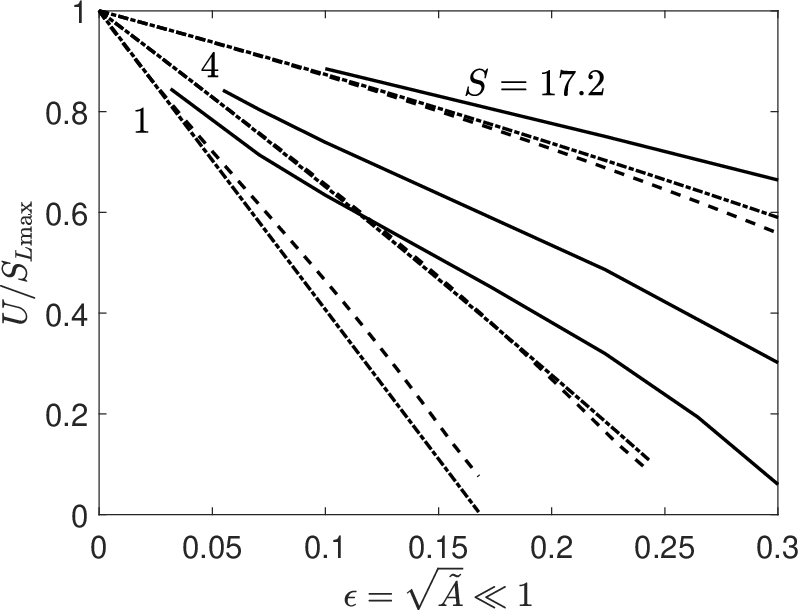}
\caption{The dashed lines are obtained from the numerical integration of~\eqref{eq:G} and the dashed-dotted curves are the results of the perturbation analysis given in the final equation in the appendix. All solid curves are the actual two-dimensional computational results. } \label{fig:Kine}
\end{figure}

Through its relationship to the mixture-fraction function, $Z(y)$, the variation of the planar laminar burning velocity with the equivalence ratio defines the function $\St_L(y)$, which will achieve its maximum value $S_m=S_{L,\rm{max}}/S_{L\infty,s}$ at a value of $y$ denoted by $y_m$. It is evident from~\eqref{eq:G} that in the limit $\ep=0$ the constant $\Ut$ cannot be less than $S_m$ since the magnitude of the square root is never less than unity, nor can it be greater than $S_m$, since then the entire pattern would propagate faster than any element of the front. Hence, at leading order in $\ep$, the pattern must propagate at the velocity $\Ut=S_m$ and the solution at this order then becomes simply
\begin{equation}
    g(y) = \pm \left(\frac{S_m^2}{\St_L^2}-1\right)^{1/2}, \label{eq:Gs}
\end{equation}
the upper sign applying for $y>y_m$ and the lower sign for $y<y_m$.

Since $\Ut=S_m$ at leading order, the first correction to $\Ut$ arising from the front curvature is determined by the variation of $\St_L(y)$ in the vicinity of the point $y=y_m$. With the normal quadratic variation about the maximum, in the first approximation $\St_L(y)=S_m-(a/2)(y-y_m)^2$, where the positive constant $a$ is the negative of the second derivative of $\St_L(y)$ at $y_m$. Then, it is found that the order $\ep$ perturbation to $g(y)$ diverged in proportion to $1/(y-y_m)$ as $y$ approaches $y_m$ unless $\Ut=S_m-\ep\M(y_m) (a/S_m)^{1/2}$. This determines the first correction to the propagation velocity, and further pursuit of the perturbation analysis, summarized in the appendix for the values $\M=\M'=1$, which apply in the thermo-diffusive approximation in the limit $\bes\rightarrow\infty$~\cite{Clavin1985}, serves to determine subsequent corrections to the location $y=y_m$ of the turning point (nose of the pattern), $y=y_t$, as well as the front shape.

Representative results from numerical integration of~\eqref{eq:G} with $\M=\M'=1$ are shown as dashed curves in the upper plots in figure~\ref{fig:reactrate}. The dashed curves are seen to coincide well with the contours of maximum reaction rate, irrespective of the value of $S$, even exhibiting a good agreement, where $\ep>0.2$ for $S=17.2$. Thin-flame descriptions therefore may be considered to be quite robust for describing the C-shaped curves in these problems.

Predictions of propagation velocities of C-flame patterns are compared in figure~\ref{fig:Kine} for $\bes=8$. As can be seen, the thin-flame propagation-velocity predictions thus are somewhat less robust than the flame-shape predictions. Although the predicted linear dependence of propagation velocities on $\ep$ is seen to be good, it decreases more rapidly with $\ep$ than is found by the full integration. This is likely a consequence of selecting a Markstein number of unity; this value of $\bes$ is small enough that the variation with $\ep$ may be expected to be weaker than would occur in the limit $\bes\rightarrow\infty$. The thin-flame description employing the expansion derived in the appendix is seen to be in good agreement with results of the numerical integration of the thin-flame equation, whence the thin-flame formulation is likely to be reasonably accurate with proper evaluations of Markstein numbers.  

\section{Conclusions}

The main conclusion to be drawn from this investigation is that not all partially premixed flames in counterflow configurations may be expected to exhibit the classical tribrachial structure of rich and lean premixed flames with a diffusion flame trailing behind. Especially at high values of the dilution-adjusted stoichiometric fuel-air ratio, such as values appropriate for methane-air flames, the diffusion flame may fade into the lean wing, with the triple flame then evolving into a fuel-rich C-shaped premixed flame, however, heat release may modify this configuration, bringing the trailing diffusion flame back into visibility, yet the asymmetry in the direction identified here is likely to remain. Totally symmetric triple flames therefore should not be anticipated to be prevalent in practical situations. 

A further notable finding is that, although not at all applicable to retreating or even to advancing edge flames, thin-flame approximations enable Markstein numbers to be applied, employing quite general chemical kinetics, to reduce the problem involving ordinary differential equations in place of more complex partial differential equations. Moreover, in the limit of small strain rates, instead of integrating ordinary differential equations, sequential solutions of purely algebraic equations suffice to produce the results that are needed. It could be worthwhile to extend simplifications of that type to address the important influences of density changes associated with the heat release.   

\section*{Acknowledgements}

This work was supported by the US AFOSR Grant No. FA9550-16-1-0443.

\section*{Appendix}

It is convenient to employ the re-scalings $\hat U = \tilde U/S_m$, $\hat S_L=\St_L/S_m$ and $\hat \ep= \ep/S_m$, before introducing the perturbation series $g=g_o+\hat\ep g_1 + \hat\ep^2 g_2 + \cdot\cdot\cdot$ and $\hat U = U_o + \hat\ep U_1 + \hat\ep^2 U_2 + \cdot\cdot\cdot$ into~\eqref{eq:G}. Then the resulting problem becomes purely algebraic in nature for the unknown quantities at each successive order. A unique choice of the eigenvalue at each order makes the solution uniformly valid in $y$ by eliminating non-analyticity at the turning point. A Taylor's expansion of $\hat S_L(y)$ around the maximum point is needed,
\begin{equation}
    \hat S_L = 1 + \delta^2 \frac{\hat S_m''}{2!} + \delta^3 \frac{\hat S_m'''}{3!} + \delta^4 \frac{\hat S_m^{iv}}{4!} + \cdot \cdot \cdot,
\end{equation}
where $\delta=y-y_m$, and $\hat S_m'', \hat S_m''',...$ are derivatives of $\hat S_L(y)$ evaluated at the maximum location. The turning-point location is also an unknown quantity that can be expanded in series $y_t = y_m + \hat\ep y_1 + \hat\ep^2 y_2 + \cdot\cdot\cdot$. 

As discussed before, the leading-order solution is
\begin{equation}
    g_o(y) = \pm \left(\frac{U_o^2}{\hat S_L^2}-1\right)^{1/2}, \quad U_o=1.
\end{equation}
where $g_o(y)$ is real ($U_o\nless 1$) and continuous ($U_o\ngtr 1$) at $\delta=0$. Collecting terms of $O(\hat\ep)$ and solving for $g_1(y)$ gives
\begin{equation}
    g_1(y) = \frac{U_1 }{\hat S_L^2 g_o} + \frac{y}{\hat S_L^2} +  \frac{1}{g_o} \frac{\dd g_o}{\dd y}.
\end{equation}
The behaviour of $g_1(y)$ as it approaches the turning point $y\rightarrow y_t= y_m + \hat\ep y_1$ is found to be $g_1(y) \sim\delta^{-1}[U_1/\sqrt{-\hat S_m''} + 1] + O(1)$, thus $U_1$ will be chosen to make $g_1(y)$ be bounded. The turning point at this order is obtained as the value of $y$ at which $g_o(y)+\hat\ep g_1(y)=0$.

In the same spirit, the equation for $g_2(y)$ is obtained at the  next order from,
\begin{align}
    g_2(y) = \frac{U_2}{\hat S_L^2 g_o} + \frac{yg_1}{\hat S_L^2 g_o} - \frac{ g_1^2}{2g_o} + \frac{\hat S_L^2 g_o g_1^2}{2}\notag\\ + \frac{1}{g_o}\frac{\dd g_1}{\dd y} - 2 \hat S_L^2g_1 \frac{\dd g_o}{\dd y} + \frac{g_o}{\hat S_L^2},
\end{align}
its behaviour as $y\rightarrow y_m + \hat\ep y_1 + \hat\ep^2 y_2$ diverges like $\delta^{-1}$ as before and $U_2$ is chosen so as to remove this divergence. The selection of the value of $U_1$ has eliminated a term of order $\delta^{-2}$, which otherwise would appear. The uniform solution at this order is given by 
\begin{equation}
g(y) =\pm\left\| \pm|g_o + \hat\ep g_1| + \hat\ep^2 g_2\right\| + O(\hat\ep^3),
\end{equation}
with the new turning point $y_t=y_m+\hat\ep y_1 + \hat\ep^2 y_2$ now being determined by requiring the expression inside the outer absolute value signs to vanish there. At this order, the eigenvalue is
\begin{equation}
    \hat U= 1 - \hat\ep \sqrt{-S_m''} + \hat \ep^2 \left(\frac{\gamma_1^2}{2} -\gamma_2 - y_m \gamma_1\right) + O(\hat\ep^3).\label{eq:perturb}
\end{equation}
where
\begin{equation}
    \gamma_1 = y_m - \frac{\hat S_m'''}{3(-\hat S_m'')}, \ \gamma_2 = 1- \frac{7\hat S_m'''^2}{72\hat S_m''^2} - \frac{\hat S_m^{iv}}{8(-\hat S_m'')}.
\end{equation}

\bibliographystyle{elsarticle-num}

\bibliography{references}



\end{document}